\documentclass[preprint,prd,amsmath,amssymb,aps,nofootinbib]{revtex4-1}

\usepackage{graphicx}
\usepackage{siunitx}
\usepackage{bm}
\usepackage{multirow}

\graphicspath{{figures/}}

\def\mathbi#1{\textbf{\em #1}}

\def\apj{Astrophys.\ J.\ }

\def\apss{Astrophys.\ Space\ Sci.\ }

\begin{document}
\title{Constraining the slow-diffusion zone size and electron injection spectral index for the Geminga pulsar halo}
\author{Kun Fang}

\email{fangkun@ihep.ac.cn}
\affiliation{
Key Laboratory of Particle Astrophysics, Institute of High Energy
Physics, Chinese Academy of Sciences, Beijing 100049, China \\
}

\date{\today}

\begin{abstract}
Measuring the electron diffusion coefficient is the most straightforward task in the study of gamma-ray pulsar halos. The updated measurements of the spatial morphology and spectrum of the Geminga halo by the High-Altitude Water Cherenkov (HAWC) experiment enable us to constrain parameters beyond the diffusion coefficient, including the size of the slow-diffusion zone and the electron injection spectrum from the pulsar wind nebulae (PWNe). Based on the two-zone diffusion model, we find that the slow-diffusion zone size ($r_*$) around Geminga is within the range of $30-70$~pc. The lower boundary of this range is determined by the goodness of fit of the model to the one-dimensional morphology of the Geminga halo. The upper limit is derived from fitting the gamma-ray spectrum of the Geminga halo, along with the expectations for the power-law index of the injection spectrum based on simulations and PWNe observations, i.e., $p\gtrsim1$. With $r_*$ set at its lower limit of $30$~pc, we obtain the maximum $p$ permitted by the HAWC spectrum measurement, with an upper limit of $2.17$ at a $3\sigma$ significance. Moreover, we find that when $r_*=30$~pc and $p=2.17$, the predicted positron spectrum generated by Geminga at Earth coincides with the AMS-02 measurement in the $50-500$~GeV range.
\end{abstract}

\maketitle

\section{Introduction}
\label{sec:intro}
According to the standard evolution model, pulsar wind nebulae (PWNe) associated with middle-aged pulsars ($t\sim100$~kyr) are typically in the bow-shock stage \cite{Gaensler:2006ua}, during which they traverse the interstellar medium (ISM) while maintaining a compact size of $\lesssim1$~pc. High-energy electrons and positrons\footnote{\textit{Electrons} will denote both electrons and positrons hereafter if not specified.} that escape from the PWNe and diffuse into the ISM generate gamma-ray halos around their respective pulsars via inverse Compton (IC) scattering of background photons. These phenomena are referred to as pulsar halos \cite{Sudoh:2019lav,Fang:2022fof,Liu:2022hqf,Lopez-Coto:2022igd}. Since the morphology of pulsar halos traces the spatial distribution of their parent electrons, they serve as ideal indicators for investigating cosmic-ray propagation in localized regions of the Galaxy \cite{Evoli:2018aza,Kun:2019sks,Liu:2019zyj,Wang:2021xph,Fang:2021qon}. Furthermore, the gamma-ray spectrum of pulsar halos is determined by the energy spectrum of the escaped electrons rather than those remaining confined within the PWNe, thus providing unique insight into the process of electron escape from the PWNe \cite{Fang:2022mdg,Fang:2022qaf}.

The diffusion coefficient inferred from the morphology of the identified pulsar halos is about two orders of magnitude smaller than the Galactic average \cite{Abeysekara:2017old,Aharonian:2021jtz,Fang:2022qaf,HAWC:2023jsq}, which also ensures that these halos are bright enough to be detected by current experiments. The slow-diffusion phenomenon around the pulsars has a crucial impact on interpreting issues such as the cosmic-ray positron excess \cite{Hooper:2017gtd,Fang:2018qco,Tang:2018wyr,Shao-Qiang:2018zla,Fang:2019ayz,Manconi:2020ipm,Wu:2022kia,Schroer:2023aoh} and the diffuse TeV gamma-ray excess \cite{Linden:2017blp,Dekker:2023six,Yan:2023hpt}. To accurately address these issues, comprehension of the slow-diffusion mechanism is essential. This slow diffusion could be attributed to a turbulent environment generated by the host supernova remnants of the pulsars \cite{Kun:2019sks} or by the escaped electrons themselves \cite{Evoli:2018aza,Mukhopadhyay:2021dyh}. Alternatively, it may be explained by the projection effects under anisotropic diffusion without needing an additional turbulent magnetic field \cite{Liu:2019zyj,DeLaTorreLuque:2022chz,Fang:2023axu}. It has also been proposed that the steep profile of pulsar halos can be explained without the diffusion coefficient suppression as long as the relativistic correction is considered in the electron propagation equation \cite{Recchia:2021kty}. However, this model provides poorer fits to the data when compared to the slow-diffusion scenario and requires an energy conversion efficiency significantly larger than $100\%$ \cite{Bao:2021hey}, which is unreasonable.

A definitive conclusion on the slow-diffusion mechanism remains elusive. Nonetheless, we can still investigate the physical parameters associated with pulsar halos phenomenologically. Recently, the High-Altitude Water Cherenkov (HAWC) experiment provided updated observations of a prototypical pulsar halo, namely the Geminga halo \cite{HAWC:2023bfh}. Compared to the initial work on the Geminga halo \cite{Abeysekara:2017old}, the precision of the spatial morphology measurement has been significantly enhanced, and the gamma-ray flux has been accurately measured in distinct energy ranges. We will show that these improvements enable us to provide valuable constraints on the size of the slow-diffusion zone and the form of the PWN electron injection spectrum, thereby broadening our scope beyond simple diffusion coefficient measurements.

In this study, we ground our calculation in the two-zone diffusion model, where the slow-diffusion zone is confined to a specific vicinity around the pulsar. This phenomenological model could appropriately characterize the electron propagation associated with pulsar halos. In Sec.~\ref{sec:2zone}, we present the calculation of the morphology and energy spectrum of the Geminga halo. In Sec.~\ref{sec:size}, we fit the latest one-dimensional morphology measured by HAWC and give a lower limit for the slow-diffusion zone size using the goodness-of-fit test. In Sec.~\ref{sec:index}, we fit the gamma-ray energy spectrum of Geminga. The HAWC measurement shows a very hard spectrum in $2-20$~TeV. We demonstrate that this feature can be interpreted by a reasonable electron injection spectrum under the two-zone diffusion model. We provide constraints on the index of the power-law term of the injection spectrum and also estimate the maximum slow-diffusion zone size. Based on these parameter constraints, Sec.~\ref{sec:positron} revisits the potential for Geminga to account for the cosmic-ray positron excess. Sec.~\ref{sec:conclu} is the conclusion.

\section{Two-zone Diffusion model}
\label{sec:2zone}
The magnetic field turbulence downstream of the SNR shock could be significantly enhanced compared to the far upstream, resulting in a downstream diffusion coefficient two orders of magnitude smaller than the ISM \cite{Kun:2019sks,2022ApJ...932...65W}. This suggests that the slow-diffusion environment can be understood if the pulsar remains within the downstream region of its corresponding SNR \cite{Kun:2019sks}. Among the observed pulsar halos, the pulsar associated with the Monogem halo indeed resides within the downstream region of its corresponding SNR, which has been strongly supported by pulsar scintillation observations \cite{Yao:2022cse}. Given the scale of SNRs, the slow-diffusion zone around pulsars may span several tens of parsecs.

Alternatively, if the slow diffusion is attributed to the escaped electrons from PWNe amplifying the surrounding magnetic turbulence via streaming instability, the diffusion coefficient could only be significantly reduced in the vicinity of pulsars as the growth rate of turbulence is directly proportional to the electron number density gradient \cite{Evoli:2018aza}. Therefore, a two-zone diffusion model with a slow-diffusion zone spanning several tens of parsecs is a plausible approximation. The current paper does not discuss the scenario of interpreting pulsar halos with anisotropic diffusion.

High-energy electrons are accelerated by PWNe and subsequently released into the ISM. The electron propagation in the ISM could be described by the diffusion-loss equation as
\begin{equation}
  \frac{\partial n(E_e, \mathbi{r}, t)}{\partial t} = \nabla \cdot[D(E_e, \mathbi{r})\nabla n(E_e, \mathbi{r}, t)] + \frac{\partial[b(E_e)n(E_e, \mathbi{r}, t)]}{\partial E_e} + Q(E_e, \mathbi{r}, t)\,,
 \label{eq:prop}
\end{equation}
where $n\equiv dN/dE_e$ is the differential electron number density, and $E_e$ is the electron energy. The diffusion coefficient in the two-zone diffusion model takes the form of 
\begin{equation}
 D(E_e, \mathbi{r})=\left\{
 \begin{aligned}
  & D_{100}(E_e/{\rm 100~TeV})^\delta, \quad & |\mathbi{r}-\mathbi{r}_p|\leq r_* \\
  & D_{100,{\rm ism}}(E_e/{\rm 100~TeV})^{\delta_{\rm ism}}, \quad & |\mathbi{r}-\mathbi{r}_p|>r_* \\
 \end{aligned}
 \right.~,
 \label{eq:2zone}
\end{equation}
where $\mathbi{r}_p$ is the location of the pulsar, $r_*$ is the size of the slow-diffusion zone, $D_{100}$ and $\delta$ are diffusion parameters for the slow-diffusion zone, and $D_{100,{\rm ism}}$ and $\delta_{\rm ism}$ are those for the typical ISM. Considering the energy range of the HAWC measurement, we set the reference energy for the diffusion coefficient at $100$~TeV. This ensures that $D_{100}$ is minimally affected by the selection of $\delta$. We adopt $D_{100,{\rm ism}}=3.4\times10^{30}$~cm~s$^{-1}$ and $\delta_{\rm ism}=0.38$ as inferred from the cosmic-ray boron-to-carbon ratio \cite{Yuan:2017ozr}.

The second term on the right-hand side of Eq.~(\ref{eq:prop}) is the energy-loss term, where $b(E_e)\equiv|dE_e/dt|$ is the electron energy-loss rate. We take the magnetic field strength of $B=3$~$\mu$G for the synchrotron loss rate. For the IC scattering, we adopt the seed photon field given in Ref.~\cite{Abeysekara:2017old} and the parametrization method given in Ref.~\cite{Fang:2020dmi} to calculate the energy-loss rate. 

The third term on the right-hand side of Eq.~(\ref{eq:prop}) is the source term. Assuming the time profile of the source function follows the pulsar spin-down luminosity, it takes the form of
\begin{equation}
 Q(E_e,\mathbi{r},t)=\left\{
 \begin{aligned}
 & q(E_e)\,[(t_p+t_ { \rm
sd})/(t+t_{\rm sd})]^2\,\delta(\mathbi{r}-\mathbi{r}_p)\,, & t\geq0 \\
 & 0\,, & t<0
 \end{aligned}
 \right.\,,
 \label{eq:src}
\end{equation}
where $q(E_e)$ is the current electron injection spectrum, $t_p$ is the characteristic age of the pulsar, $t_{\rm sd}$ is the pulsar spin-down timescale, and $t=0$ corresponds to the birth time of the pulsar. The characteristic age of Geminga is $t_p=342$~kyr as given by the Australia Telescope Nation Facility (ATNF) pulsar catalog \cite{Manchester:2004bp}, and $t_{\rm sd}$ is set to be $10$~kyr. 

The injection energy spectrum is assumed to be a power law with a super-exponential cutoff as
\begin{equation}
 q(E_e)\propto E_e^{-p}\,{\rm exp}\left[-\left(\frac{E_e}{E_c}\right)^2\right]\,,
 \label{eq:inj}
\end{equation}
where the form of the cutoff term is suggested by the mechanism of the relativistic shock acceleration \cite{Dempsey:2007ng}. 

We numerically solve Eq.~(\ref{eq:prop}) using the finite volume method \cite{Fang:2018qco}. A semianalytical method is also applicable for two-zone diffusion \cite{Osipov:2020lty}. We integrate the resulting electron number density $n$ over the line of sight from Earth to the vicinity of the pulsar to obtain the electron surface density. This allows us to derive the one-dimensional gamma-ray morphology of the halo, $s(\theta,E_\gamma)$, using the standard calculation of IC scattering \cite{Blumenthal:1970gc}, where $\theta$ is the angle away from the pulsar. The gamma-ray spectrum within an arbitrary angular radius around the pulsar is then calculated by 
\begin{equation}
 \frac{dF}{dE_\gamma}(\theta<\theta_0,E_\gamma)=\int_0^{\theta_0} s(\theta,E_\gamma)2\pi\theta d\theta~.
 \label{eq:spec}
\end{equation}
The distance of the Geminga pulsar to the Earth takes $250$~pc as measured by trigonometric parallax \cite{2007Ap&SS.308..225F}. 

\section{Lower limit of the slow-diffusion zone size}
\label{sec:size}
The one-dimensional gamma-ray morphology of the Geminga halo provided by HAWC, also known as the surface brightness profile (SBP), exhibits a median energy of $\approx20$~TeV \cite{HAWC:2023bfh}. This corresponds to a parent electron energy of $\approx100$~TeV \cite{Abeysekara:2017old}. Therefore, the shape of the SBP is predominantly determined by the diffusion coefficient at $100$~TeV and is minimally influenced by parameters describing energy dependence, such as $\delta$ or $p$. In the fitting process of this section, we consider $D_{100}$ and the normalization as free parameters. The values of $\delta$ and $p$ remain fixed as in the initial HAWC paper \cite{Abeysekara:2017old}, and we find that variations in these parameters have negligible impact on the fitting results. To compare with the HAWC data, we also convolve a point spread function with a $68\%$ containment radius of $0.3^\circ$ on the model.

The SBP centered on Geminga provided by HAWC includes contributions from both the Geminga and Monogem halos. The component of the Monogem halo is determined using the SBP measurement center on Monogem, where only the data within a radius of $3^\circ$ is considered to avoid significant influence from the Geminga halo at larger angles. The Monogem halo is characterized using a one-zone diffusion model, yielding $D_{100}=4.6\times10^{27}$~cm~s$^{-1}$. We fix the Monogem halo component in the following fitting process.

\begin{figure}[t!]
\centering
\includegraphics[width=0.49\textwidth]{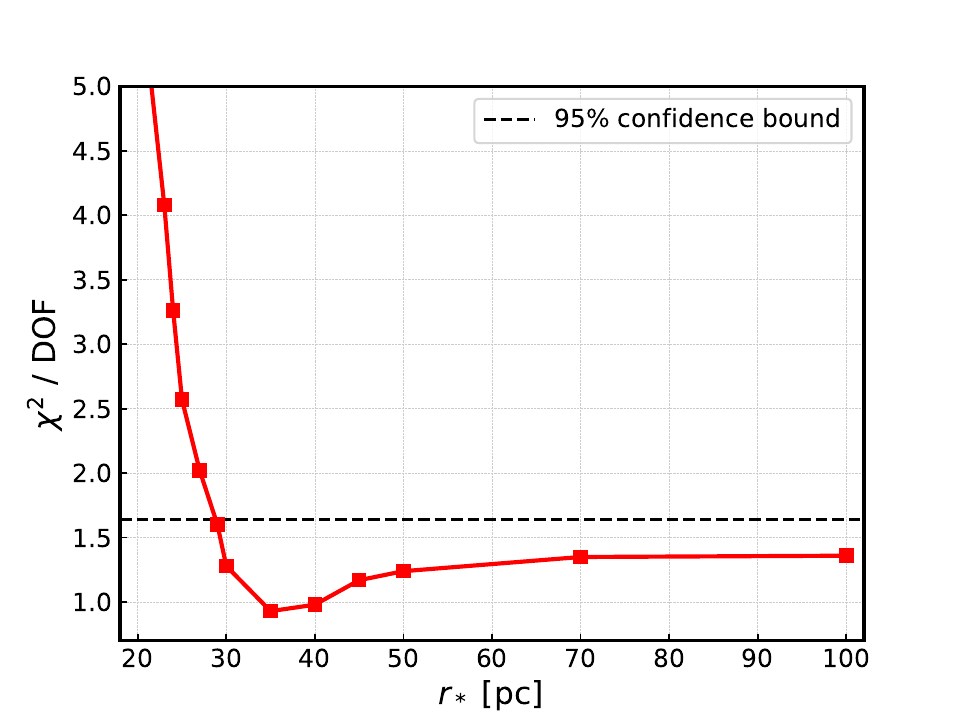}
\includegraphics[width=0.49\textwidth]{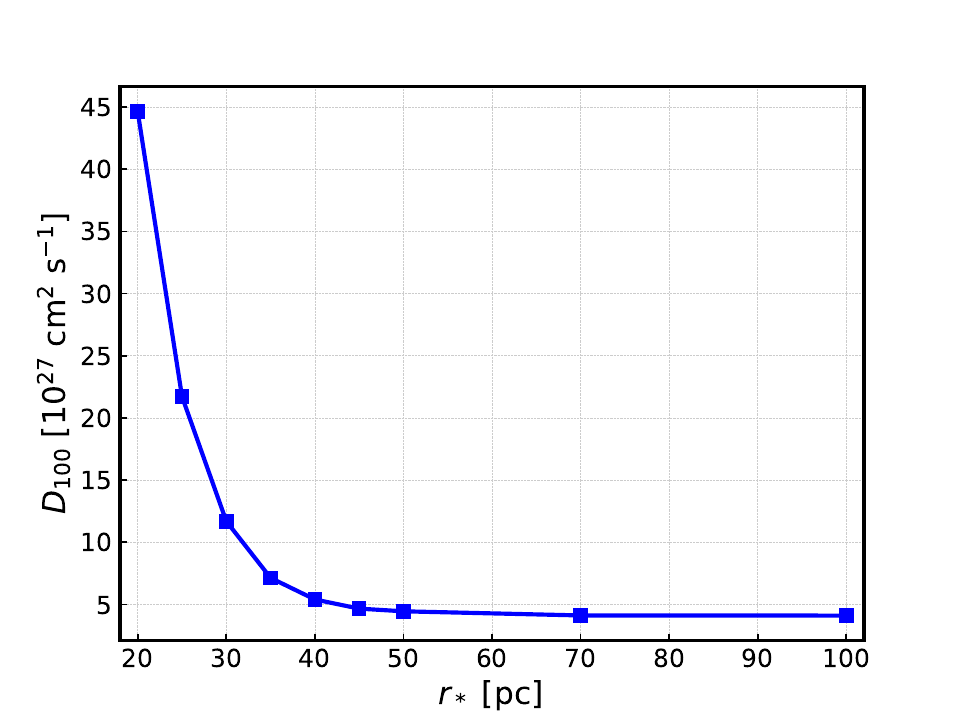}
\caption{The fitting results for the Geminga halo SBP measured by HAWC assuming different slow-diffusion zone sizes ($r_*$). The left panel shows the minimum reduced chi-square statistic with 16 d.o.f. The right panel depicts the corresponding diffusion coefficient of the slow-diffusion zone at the energy of $100$~TeV.}
\label{fig:size_chi2}
\end{figure}

We assume varying sizes for the slow-diffusion zone and fit these models to the SBP data centered on Geminga. The minimum reduced chi-square statistic $\chi^2$ for each $r_*$ is illustrated in the left panel of Fig.~\ref{fig:size_chi2}. With $16$ degrees of freedom (d.o.f.), the $95\%$ confidence bound for the reduced $\chi^2$ is $1.64$, as marked by the dashed line in the figure. It can be seen that models with $r_*\lesssim30$~pc are excluded at the $95\%$ confidence level by the chi-square test. The chi-square statistic is minimized at $\approx1.0$ when $r_*=35$~pc. Beyond $r_*\gtrsim50$~pc, there is no significant change in the goodness of fit. The corresponding $D_{100}$ for each $r_*$ is depicted in the right panel of Fig.~\ref{fig:size_chi2}. When $r_*\gtrsim40$~pc, the best-fit $D_{100}$ tends to be $\approx4.5\times10^{27}$~cm$^2$~s$^{-1}$, consistent with that given by the HAWC paper. For models with $r_*<40$~pc, a larger $D_{100}$ is required to widen the SBP to match the data.

\begin{figure}[t!]
\centering
\includegraphics[width=0.6\textwidth]{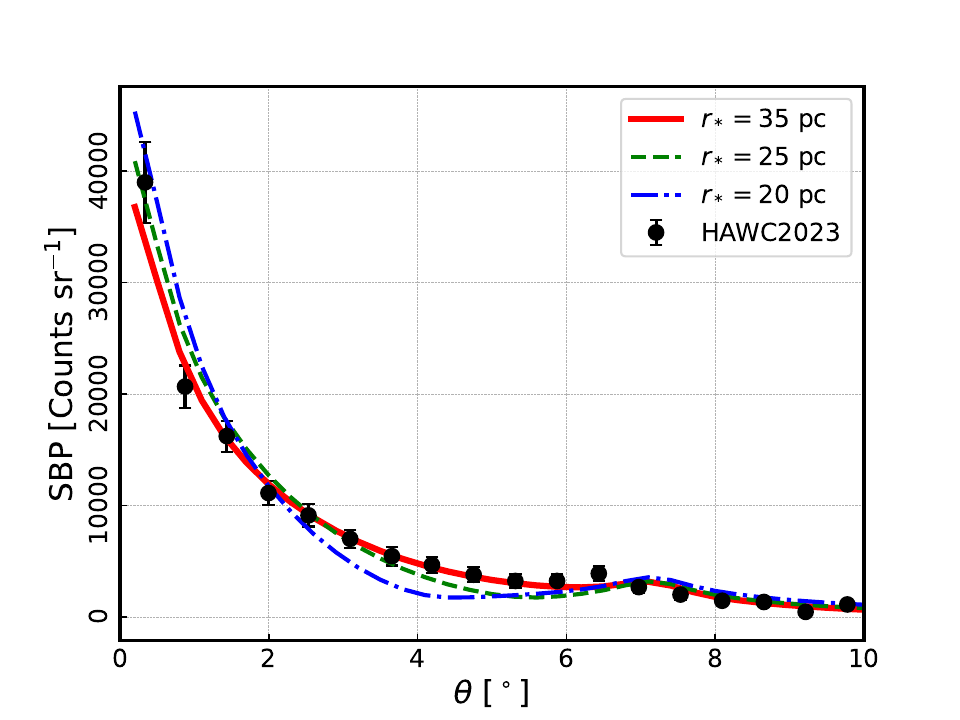}
\caption{Best-fit models for the SBP centered on Geminga under different assumptions of the slow-diffusion zone size. This figure serves to explain the exclusion of models with a small $r_*$ intuitively. The case of $r_*=35$~pc provides the best fit among all models, with $\chi^2/{\rm d.o.f.}=0.93$, as shown in the left panel of Fig.~\ref{fig:size_chi2}.}
\label{fig:size_prof}
\end{figure}

Figure~\ref{fig:size_prof} provides an intuitive explanation for excluding models with a small $r_*$. In the two-zone diffusion model, electrons disperse rapidly once they leave the slow-diffusion zone, resulting in a significant flux reduction in the fast-diffusion zone compared to the slow-diffusion zone. The flux of the fast-diffusion zone is so low that it is difficult to distinguish them from the background. As a result, the SBP of the pulsar halo under two-zone diffusion appears to be more contracted if $r_*$ is smaller. As shown in the right panel of Fig.~\ref{fig:size_chi2}, the best-fit value of $D_{100}$ determined by the fitting process increases in an attempt to compensate for the reduction in the halo extent caused by the decrease in $r_*$. If $r_*$ is too small, a significant reduction in flux will occur at a location too close to the pulsar. For instance, in the case of $r_*=20$~pc depicted in Fig.~\ref{fig:size_prof}, the model significantly underestimates the data in the angular range of $\approx2^\circ$ to $6^\circ$ due to the transition from the slow- to fast-diffusion zone. This results in a very poor goodness of fit that cannot be rectified by assuming a larger $D_{100}$. 

On the other hand, the current SBP measurement does not constrain the upper limit of $r_*$. For Geminga, an angular distance of $10^\circ$ corresponds to a transverse spatial distance of $\approx45$~pc. This means that when $r_*$ exceeds $45$~pc, the variation in SBP within $10^\circ$ predicted by different models is insignificant. However, as we will show in the next section, the gamma-ray energy spectrum could enable us to estimate the upper limit of $r_*$.

Although HAWC also provides the preliminary SBP centered on Monogem, since the Monogem halo is significantly dimmer than the Geminga halo, the SBP is heavily influenced by the Geminga halo at large angles. The Geminga halo may display asymmetry in its morphology \cite{LHAASO:2023chen}, but in the absence of two-dimensional data, we cannot estimate the contribution of the Geminga halo to this SBP accurately. Therefore, we do not investigate the slow-diffusion zone size corresponding to the Monogem halo in this study.

\section{Constraints of gamma-ray spectrum on parameters}
\label{sec:index}
The HAWC spectrum measurement of the Geminga halo ranging from $\approx1-100$~TeV indicates that the electron injection spectrum is consistent with the form presented in Eq.~(\ref{eq:inj}), with $E_c\approx100$~TeV. Nevertheless, the gamma-ray spectrum from $2$~TeV to $20$~TeV exhibits a hard nature, indicating the necessity for a hard power-law term in the injection spectrum. However, under one-zone diffusion, even though a very hard injection spectrum of $p\approx1$ is assumed, the model still does not fit the observed gamma-ray spectrum well \cite{HAWC:2023bfh}.

Assuming the one-zone diffusion model with $D_{100}=5\times10^{27}$~cm$^2$~s$^{-1}$ and $\delta=1/3$, as used in the HAWC work, the gamma-ray flux within $10^\circ$ around the pulsar constitutes $\approx98\%$ of the total angle-integrated spectrum. Consequently, the spectral measurement provided by HAWC approximately represents the integrated gamma-ray spectrum within $\theta=10^\circ$.

\begin{figure}[t!]
\centering
\includegraphics[width=0.6\textwidth]{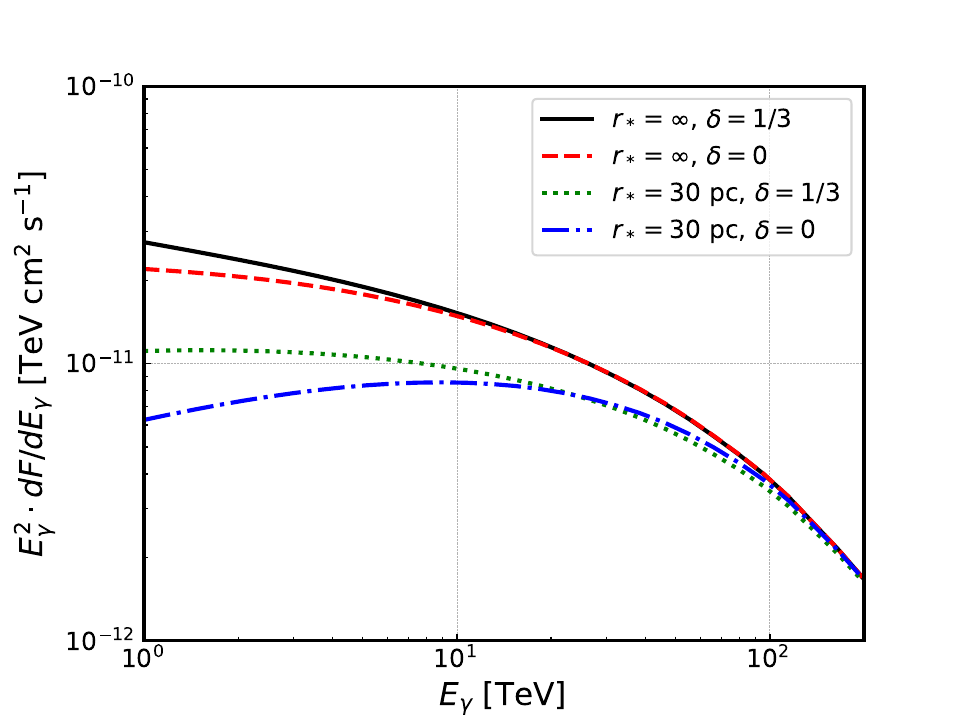}
\caption{This figure illustrates the impact of the energy index of the diffusion coefficient ($\delta$) and the size of the slow-diffusion zone ($r_*$) on the angle-integrated gamma-ray spectrum within $10^\circ$ around the Geminga pulsar. The electron injection spectrum is assumed to be a power law with $p=2.2$. The gamma-ray spectral softening at high energies is due to the Klein-Nishina effect.}
\label{fig:spec_display}
\end{figure}

The integrated gamma-ray spectrum within a specific angle $\theta_0$ is determined not only by the electron injection spectrum but also by $\delta$ and $r_*$, as illustrated in Fig.~\ref{fig:spec_display}, where $\theta_0=10^\circ$. Given that $D_{100}$ is constrained by the SBP data, a smaller $\delta$ corresponds to faster diffusion of low-energy electrons, making them more likely to escape from $\theta_0$ and consequently yielding a harder gamma-ray spectrum. Moreover, within the energy range of interest, the characteristic diffusion distance ($\sim\sqrt{D\tau}$, where $\tau$ is the smaller one between the electron cooling time and the pulsar age) of low-energy electrons is larger than that of high-energy electrons. This implies that a smaller $r_*$ results in a higher proportion of low-energy electrons escaping from $\theta_0$, thereby producing a harder gamma-ray spectrum. As can be seen from Fig.~\ref{fig:spec_display}, the impact of $r_*$ on the spectral shape is significant. Therefore, under the two-zone diffusion model, a very hard injection spectrum might not be necessary. This point has also been demonstrated in our previous studies \cite{Fang:2021qon,Fang:2022qaf}.

The analysis in Sec.~\ref{sec:size} indicates that $r_*=30$~pc may serve as the lower limit for the slow-diffusion size, and we typically regard $\delta=0$ as the lower limit for the energy index of the diffusion coefficient. Such energy-independent diffusion of electrons can occur if their transport is governed by the field-line random walk within the ISM turbulence \cite{Reichherzer:2021yyd}, or when the electrons are scattered by fast-mode turbulence in a regime where the damping of the turbulence cascade is collisionless \cite{Yan:2007uc}. With this set of parameters, we explore the maximum permissible injection spectral index derived from the observation. We fit the integrated gamma-ray spectrum obtained from the model within $\theta=10^\circ$ to the spectral measurements of the Geminga halo by HAWC (including the flux upper limits with $95\%$ confidence), with $p$, $E_c$, and the normalization as free parameters. We impose constraints on $p$ using the likelihood ratio test. The likelihood ratio is defined as $\lambda=-2\ln\left[L(p_0)/L(p)\right]$, where $L(p)$ is the global maximum likelihood, and $L(p_0)$ is the local maximum likelihood when the spectral index is fixed at $p_0$. According to Wilks' theorem, $\lambda$ follows the chi-square distribution with one degree of freedom. When $\lambda$ equals $4$ and $9$, we can respectively determine the $2\sigma$ and $3\sigma$ confidence bounds for $p$.

\begin{figure}[t!]
\centering
\includegraphics[width=0.6\textwidth]{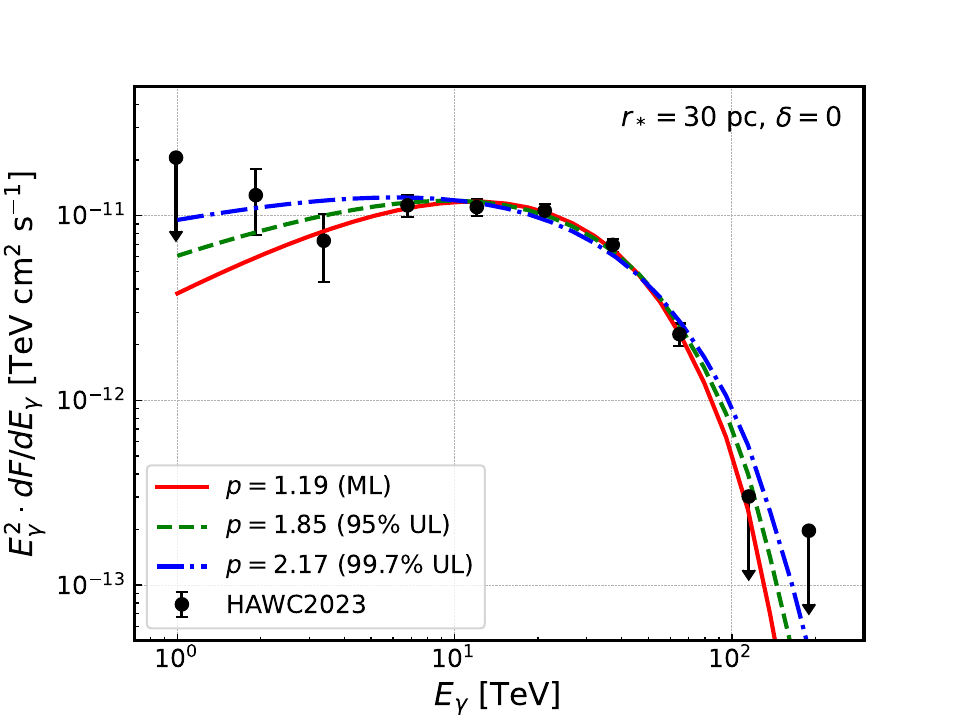}
\caption{Fitting results of the Geminga halo gamma-ray spectrum by the two-zone diffusion model. The model represents the angle-integrated spectrum within $10^\circ$ around the pulsar, adopting the smallest $r_*$ and $\delta$ to achieve the largest possible injection spectral index $p$. The three models in the figure include the maximum likelihood model and models using the $2\sigma$ and $3\sigma$ confidence upper limits of $p$. The best-fit $E_c$ for these scenarios are all around $100$~TeV.}
\label{fig:spec}
\end{figure}

The maximum likelihood $p$ we obtained is $1.18$, and the upper limits at $2\sigma$ and $3\sigma$ confidence levels are $1.85$ and $2.17$, respectively. The fitting effects of these three scenarios on the data are illustrated in Fig.~\ref{fig:spec}. It can be seen that when $p$ is larger, the model tends to overestimate the flux at low- and high-energy ranges while underestimating the flux at the intermediate-energy range. The best-fit $E_c$ in different scenarios are all around $100$~TeV. For larger $\delta$ values, such as $\delta=1/3$ or $\delta=1/2$, as predicted by Komolgorov's or Kraichnan's theory, the resulting injection spectral index is correspondingly smaller. We have reiterated the fitting procedure and summarized the results in Table~\ref{tab:rstar_30}.

\begin{table}[t!]
\caption{Power-law index of the electron injection spectrum ($p$) resulted from the fitting of the HAWC gamma-ray spectrum, under the assumption that the slow-diffusion zone size is $30$~pc. Values lower than expectations, namely $p<1$, are denoted in italics.}
\begin{ruledtabular}
\begin{tabular}{cccc}
 \multirow{2}*{$\delta$} & \multicolumn{3}{c}{$p$} \\
 \cline{2-4}
 & Maximum likelihood & $2\sigma$ upper limit & $3\sigma$ upper limit \\
 \hline
 $0$ & $1.19$ & $1.85$ & $2.17$ \\
 $1/3$ & $\textit{0.81}$ & $1.36$ & $1.82$ \\
 $1/2$ & $\textit{0.59}$ & $1.17$ & $1.65$ \\
\end{tabular}
\end{ruledtabular}
\label{tab:rstar_30}
\end{table}

On the other hand, we fit the spectral data under the assumption of varying $r_*$ values while maintaining $\delta=0$. The derived 
$3\sigma$ confidence upper limits of $p$ from these fits are summarized in Table~\ref{tab:var_rstar}, indicating a decrease in the upper limit of $p$ as $r_*$ increases. Numerical simulations suggest that the spectral index of electrons within a bow-shock PWN could approach $1$ at the colliding flow between the stellar-wind termination shock and the bow shock \cite{Bykov:2017xpo}. However, existing theories or observations do not support spectral indices to be smaller. Moreover, the x-ray PWN of Geminga exhibits two components with distinct spectral indices, with the harder and softer components corresponding to $p=1.08$ and $2.26$, respectively \cite{Posselt:2016lot}. As discussed in Ref.~\cite{Fang:2022mdg}, these components may indicate two disparate modes of electron escape, suggesting that the total injection spectrum could be a composite of both components. Consequently, for the Geminga halo, an injection spectral index of $1$ can be regarded as the lower limit. This allows us to estimate the upper limit of $r_*$. At $r_*=70$~pc, the upper limit of $p$ falls below $1$, suggesting that models with $r_*\geq70$~pc are disfavored. Combined with the lower limit obtained in Sec.~\ref{sec:size}, we are able to provide a comprehensive constraint on the slow-diffusion zone size, which lies between $30$ and $70$~pc.

\begin{table}[t!]
\caption{$3\sigma$ confidence upper limits of the power-law index of the electron injection spectrum ($p$) resulted from fitting the HAWC gamma-ray spectrum, assuming $\delta=0$. Values lower than expectations, namely $p<1$, are denoted in italics.}
\begin{ruledtabular}
\begin{tabular}{cccccc}
$r_*$ [pc] & $30$ & $40$ & $50$ & $60$ & $70$ \\
\hline
UL for $p$ & $2.17$ & $1.60$ & $1.32$ & $1.03$ & $\textit{0.87}$ \\
\end{tabular}
\end{ruledtabular}
\label{tab:var_rstar}
\end{table}

In comparison, Ref.~\cite{Schroer:2023aoh} estimates $r_*$ for the Geminga halo to be $20-50$~pc, but these constraints are not derived from a direct fit to the data.  Using the new data from HAWC, our analysis demonstrates that scenarios with $r_* < 30$~pc can be excluded with high significance, providing a more stringent lower limit for $r_*$ than Ref.~\cite{Schroer:2023aoh}. The upper limit of $50$~pc suggested by Ref.~\cite{Schroer:2023aoh} is derived from a general estimate for this type of source \cite{Hooper:2017gtd}, whereas our upper limit is based on observational constraints specific to the Geminga halo. Additionally, while Ref.~\cite{Schroer:2023aoh} argues that the power index $p$ of the electron injection spectrum is around $2.0$, our findings suggest that the uncertainty in $r_*$ necessitates a range of possible values for $p$, resulting in a more comprehensive conclusion.

Recently, the x-ray observation by eROSITA provides an upper limit of $1.4$~$\mu$G for the ISM magnetic field within $1^\circ$ around the Geminga pulsar \cite{Khokhriakova:2023rqh}. This is lower than the $3$~$\mu$G adopted in the initial HAWC study and our above calculations. Thus, we also consider a scenario with $B=1$~$\mu$G, where the synchrotron energy-loss rate for electrons is significantly reduced. This change notably affects the lifetime of higher-energy electrons, thereby influencing the estimation of the cutoff term in the electron injection spectrum. When $B$ shifts from $3$~$\mu$G to $1$~$\mu$G in the case of $r_*=30$~pc and $\delta=0$, the best-fit $E_c$ varies from $94$~TeV to $83$~TeV. However, this adjustment in magnetic field strength has a negligible effect on $p$, with its $3\sigma$ confidence upper limit merely changing from $2.17$ to $2.19$, leaving our conclusion unaffected.

\section{Implication on the positron excess}
\label{sec:positron}
The positron excess is one of the most intriguing phenomena in cosmic-ray studies. As measured by space experiments \cite{PAMELA:2008gwm,Fermi-LAT:2011baq,AMS:2013fma}, the positron spectrum above several tens of GeV significantly exceeds the predicted secondary positron flux produced by cosmic-ray nuclei spallation during Galactic propagation. Among astrophysical accelerators, pulsars (or their PWNe) are the most plausible sources of this excess \cite{Hooper:2008kg,Yuksel:2008rf,Yin:2013vaa}, given their capacity to generate high-energy electron/positron pairs.

The positron spectrum measured by AMS-02 reveals a high-energy cutoff at $\approx800$~GeV with a significance of $4.7\sigma$ \cite{Aguilar:2019owu,Kounine:2023AV}. As discussed in our previous studies, this cutoff is likely attributed to the cooling effect of the positrons generated by a dominant source rather than the superposition of multiple sources \cite{Fang:2019ayz}. This is due to the fact that pulsars or PWNe are unlikely to share a common acceleration limit at $\sim1$~TeV. Among observed pulsars, the most plausible dominant sources of the positron excess are Geminga, Monogem, and PSR B1055-52 \cite{Fang:2019ayz,Bykov:2019mis,Martin:2022hrx}, all of which are nearby and middle-aged pulsars.

Within the framework of the two-zone diffusion model, the contribution of nearby pulsars to the positron spectrum at Earth is primarily affected by parameters such as the diffusion coefficient, the slow-diffusion zone size, and the positron injection spectrum.  We use parameters determined from fitting the HAWC data (as detailed in Secs.~\ref{sec:size} and \ref{sec:index}) to predict the contribution of Geminga to the positron spectrum at Earth. As shown in Fig.~2 of Ref.~\cite{Fang:2019ayz}, the positron spectrum produced by Geminga, under assumptions of $r_*=50$~pc and $p=1.8$, is too hard compared to the AMS-02 measurement. Reducing $r_*$ or increasing $p$ can yield a softer positron spectrum. Intriguingly, as indicated in Table~\ref{tab:var_rstar}, an inverse correlation exists between the required $p$ and $r_*$ when interpreting the gamma-ray spectrum of the Geminga halo. Consequently, by setting $r_*$ to its lower limit of $30$~pc, we can achieve the softest possible positron spectrum at Earth. Furthermore, a smaller $\delta$ corresponds to faster diffusion of low-energy positrons, enabling more low-energy positrons to reach Earth and further soften the spectrum.

\begin{figure}[t!]
\centering
\includegraphics[width=0.6\textwidth]{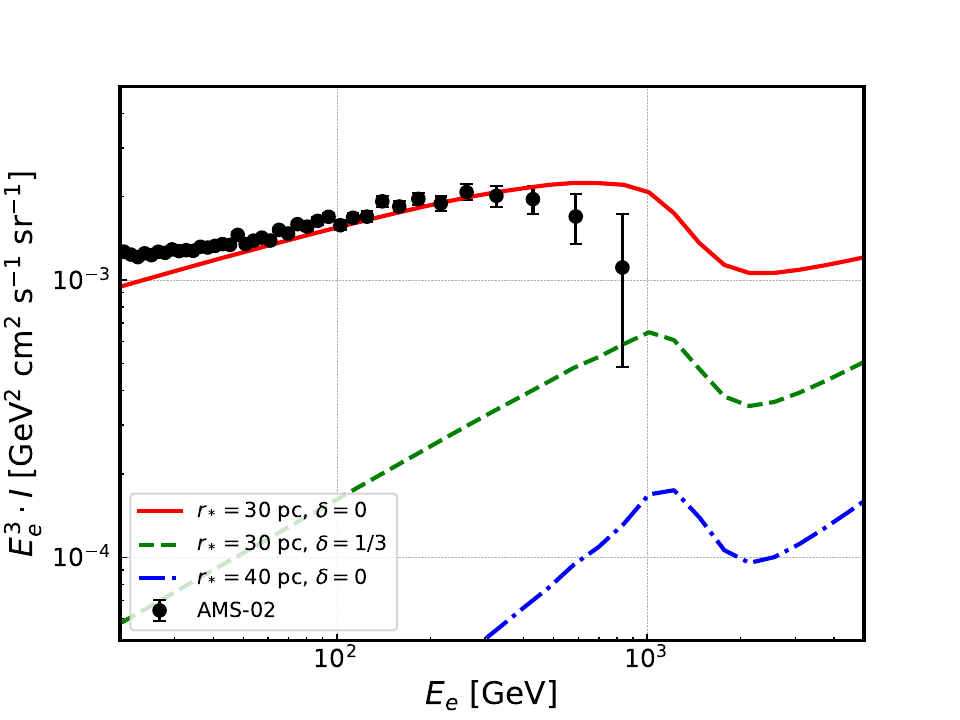}
\caption{The predicted positron spectrum from Geminga under different assumptions of $r_*$ and $\delta$, along with the corresponding $D_0$, $p$, and normalization determined by the fits in Secs.~\ref{sec:size} and \ref{sec:index}, where $p$ is adopted as the $3\sigma$ upper limits as presented in Tables~\ref{tab:rstar_30} and \ref{tab:var_rstar}. The scenario with $r_*=30$~pc and $\delta=0$ provides the most optimistic prediction to the AMS-02 measurements among all the models consistent with the HAWC measurements of the Geminga halo.}
\label{fig:posi}
\end{figure}

As an optimistic estimate, we adopt $r_*=30$~pc and $\delta=0$, along with the upper limit of the injection spectrum index $p=2.17$ and the corresponding best-fit normalization derived in Sec.~\ref{sec:index}. The predicted positron energy spectrum, $I(E_e)=n(E_e)c/4\pi$, depicted by the red solid line in Fig.~\ref{fig:posi}, is consistent with the measurements of both absolute flux and spectral shape below $\approx500$~GeV. Below $\approx50$~GeV, the secondary component may dominate the positron spectrum. However, the predicted spectral cutoff occurs at a higher energy than observed. One possible explanation is that the positrons traverse regions with higher magnetic field strength than assumed, resulting in a spectral cutoff at lower energies. Alternatively, an older pulsar, such as PSR B1055-52, could be the main contributor to the high-energy positrons instead of Geminga. Figure~\ref{fig:posi} also shows two additional scenarios with a larger $r_*$ or a larger $\delta$, where $p$ is set as the corresponding $3\sigma$ confidence limits presented in Tables \ref{tab:rstar_30} and \ref{tab:var_rstar}. Under these conditions, the positron flux produced by Geminga is insufficient to interpret the observation.

It should be noted that the calculations above are based on extrapolating the parameters derived by the HAWC measurements to lower energies. As the positron excess primarily occurs between $10$~GeV to $1$~TeV, the GeV gamma-ray measurements of the Geminga halo would impose more direct constraints on the positron injection spectrum and the diffusion coefficient within this energy range \cite{Shao-Qiang:2018zla,DiMauro:2019yvh,Zhou:2022jzg}. However, there is no definitive conclusion on the GeV measurements of the Geminga halo at present.

\section{Conclusion}
\label{sec:conclu}
Based on the latest SBP and gamma-ray spectrum measurements of the Geminga halo by HAWC, we have studied the slow-diffusion zone size ($r_*$) around Geminga and the parameters of the electron injection spectrum. Current theories expect slow diffusion to be confined within tens of parsecs around Geminga, leading us to employ a two-zone diffusion model to characterize electron propagation. Our results suggest that $r_*$ is within the range of $30-70$~pc. The lower end of this range is constrained by the goodness of fit when fitting the SBP with the model. Models with $r_*\lesssim30$~pc are disfavored by the chi-square test with a confidence level higher than $95\%$. The SBP derived from these models declines too quickly with increasing $\theta$ compared to the data, owing to a significant drop in the electron number density beyond $r_*$.

The upper limit of $r_*$ is estimated by fitting the gamma-ray spectrum of the Geminga halo, along with the expectations for the power-law index ($p$) of the electron injection spectrum based on simulations and PWN observations, where $p\gtrsim1$. Under the two-zone diffusion model, the derived $r_*$ exhibits an inverse correlation with $p$ when fitting the angle-integrated gamma spectrum within $10^\circ$ around the pulsar. For $r_*\gtrsim70$~pc, the upper limit of $p$ at a $3\sigma$ confidence level falls below $1$, implying that those models are disfavored.

Conversely, when $r_*$ is set to its lower limit of $30$~pc, we obtain the maximum $p$ permitted by the HAWC spectrum measurement, with an upper limit of $2.17$ at a $3\sigma$ confidence level. We also find that when $r_*=30$~pc and $p=2.17$, the positron flux generated by Geminga at Earth fits well with the positron spectrum measured by AMS-02 in the $50-500$~GeV range, in terms of both the spectral shape and the absolute flux. Notably, the upper limit of $p$ coincides with that required to interpret the diffuse gamma-ray excess using gamma-ray pulsar halos, where $p\approx2.2$ \cite{Yan:2023hpt}. Besides, the cutoff energy of the electron injection spectrum is determined to be $\approx100$~TeV.

In interpreting the gamma-ray spectrum of the halo, the energy index of the diffusion coefficient ($\delta$) also exhibits degeneracy with $p$ and $r_*$. We have considered several typical values for $\delta$ ($0$, $1/3$, and $1/2$). The above upper limits for $r_*$ and $p$ are derived under the condition of $\delta=0$. If $\delta$ increases, these constraints would be stricter. For example, when $\delta=1/2$, the upper limit for $p$ would decrease to $1.65$. In the near future, the Large High Altitude Air Shower Observatory (LHAASO) is expected to provide energy-dependent morphology measurements for the Geminga halo, which could break the degeneracy among the parameters.

The recent H.E.S.S. measurement to the Geminga halo has revealed an unexpected soft gamma-ray spectrum in the $1-10$~TeV range within a $1^\circ$ radius from the pulsar \cite{HESS:2023sbf}. This finding appears to be at odds with the results obtained by HAWC. The spectral features below $10$~TeV need further verification through the forthcoming LHAASO result.

\begin{acknowledgments}
The author would like to thank H. Zhou and R. Torres for sharing the energy range of the Geminga halo morphology measurements and the point spread function size of HAWC. This work is supported by the National Natural Science Foundation of China under Grants No. 12105292 and No. U2031110.
\end{acknowledgments}

\bibliography{references}

\end{document}